\documentclass[11pt]{article}
\usepackage{amssymb,amsmath,amsfonts}
\usepackage{graphicx}
\usepackage{graphics}
\usepackage{eepic,epsfig}
\usepackage{dcolumn}


\begin{document}

\title{Vacuum polarization around cosmic strings in de Sitter spacetime}
\author{A. A. Saharian \vspace{0.3cm} \\
\textit{Institute of Physics, Yerevan State University, }\\
\textit{1 Alex Manogian Street, 0025 Yerevan, Armenia }}
\date{\vspace{-5ex}}
\maketitle

\begin{abstract}
Cosmic strings are the most popular topological defects arising from the
spontaneous breaking of fundamental symmetries in the early Universe. They
are the source of a number of interesting effects in cosmology and
astrophysics. An interesting effect is the polarization of the vacuum state
for quantum fields induced by the non-trivial topology of the spacetime
around cosmic strings. In the present paper we describe the combined effects
of the straight cosmic string and the gravitational field on the local
properties of scalar, spinor and electromagnetic vacua. The local de Sitter
spacetime is considered as the background geometry. It is shown that the
influence of the spacetime curvature is essential at distances from the
string of the order of or larger than the curvature radius. A qualitatively
new feature is the appearance of the vacuum energy flux in the radial
direction with respect to the cosmic string.
\end{abstract}

\section{Introduction}

One of the most intriguing consequences of symmetry-breaking phase
transitions in the expanding early universe is the formation of topological
defects \cite{Kibb76,Vile94}. The specific type of symmetry that is broken
determines whether monopoles, cosmic strings, or domain walls are formed.
From the perspective of their physical manifestations in the current phase
of the universe expansion, cosmic strings (CS) are of special interest. The
physical phenomena that result from the influence of these defects include
gravitational lensing, the emission of high-energy cosmic rays, and the
generation of gravitational waves. In the 1980s, the CSs were also a topic
of considerable discussion as a potential source of large-scale cosmic
structure formation, offering an alternative to the mechanism proposed
within the inflationary scenario. However, observational data on the
temperature anisotropies of the cosmic microwave background radiation (CMBR)
showed that the CSs cannot be the main source of the inhomogeneities that
explain the large-scale structure of the universe, although they may have
some influence on the physical characteristics of CMBR.

In a simple model of a straight CS in the background of Minkowski spacetime,
the spacetime geometry outside the string core remains flat, albeit with a
planar angular deficit determined by the linear energy density. The
nontrivial topology of space induced by the angle deficit is a source of a
number of quantum effects. In particular, the physical characteristics of
the vacuum state for quantum fields are modified. As such characteristics,
the vacuum expectation values (VEVs) of various bilinear combinations of the
field operators were considered in the literature (see, for example, the
references given in \cite{Beze06b,Beze22T}). They include the mean field
squared, the VEVs of the energy-momentum tensor (EMT) and current density.
The main part of investigations consider the geometry of CS in the
background of Minkowski spacetime. From the point of view of cosmological
and astrophysical applications the corresponding results have to be
generalized to curved background geometries. In the present paper we review
the results of the investigations for the vacuum polarization in the
geometry of CS in the de Sitter (dS) spacetime. dS spacetime is maximally
symmetric and that allows to provide exact results for the string induced
physical characteristics of the vacuum state. These results shed light on
the influence of the gravitational field for more complicated background
geometries. Another motivation for the choice of the dS bulk is related to
its important role in cosmology. In particular, the main part of the models
of inflation in the early Universe employ the dS expansion to give natural
solutions for a number of problems in standard cosmology. The solutions of
the Einstein's equations describing the gravitational field around CS in the
dS background were considered in \cite{Line86}-\cite{Brih08}. The combined
effects of the CS and gravitational field on the local properties of the
scalar, spinor and electromagnetic vacua in the corresponding geometry have
been studied in \cite{Beze09}-\cite{Sant24b}.

The organization of the paper is as follows. In the next section we specify
the background geometry and present the results of the investigations for
the VEVs of the scalar field squared and EMT. The polarization of the
fermionic vacuum is discussed in section \ref{sec:Dirac}. The fermion
condensate and the VEV of the corresponding EMT are studied. The VEVs of the
electric and magnetic field squared and EMT for the electromagnetic field
are discussed in section \ref{sec:Elmag}.

\section{Background geometry and scalar vacuum polarization}

\label{sec:Scalar}

In this section we describe the background geometry and consider the
polarization of the scalar vacuum around CS.

\subsection{Background geometry}

We consider a spacetime geometry described by the line element 
\begin{equation}
ds^{2}=dt^{2}-e^{2t/a}(dr^{2}+r^{2}d\phi ^{2}+dz^{2}),  \label{ds}
\end{equation}%
with spatial cylindrical coordinates $(r,\phi ,z)$ varying in the range $%
0\leq r<\infty $, $0\leq \phi \leq \phi _{0}$, and $-\infty <z<+\infty $.
For the variation range of the synchronous time coordinate one has $-\infty
<t<+\infty $. In the special case $\phi _{0}=2\pi $, (\ref{ds}) corresponds
to the dS spacetime covered by inflationary coordinates. The spacetime
curvature radius is determined by the parameter $a$ expressed in terms of
the cosmological constant $\Lambda $ by the relation $a=\sqrt{3/\Lambda }$.
For $\phi _{0}<2\pi $ the line element (\ref{ds}) presents the geometry
around an idealized CS with zero thickness core and aligned along the $z$%
-axis. The planar angle deficit is determined by the energy $\mu _{0}$ of
the CS per unit length by the relation $2\pi -\phi _{0}=8\pi G\mu _{0}$,
where $G$ is the Newton gravitational constant. In this simplified model,
the presence of the CS does not change the local geometry in the region $r>0$%
. However, the spatial topology is changed and this is a source of a number
of interesting effects. Our main concern here is the influence of the
nontrivial topology on the local characteristics of the vacuum state for
scalar, spinor and electromagnetic fields. In the region $r>0$ the line
element (\ref{ds}) is conformally flat. That is seen introducing the
conformal time coordinate $\tau =-ae^{-t/a}$ with the range of variation $%
-\infty <\ \tau \ <\ 0$. The line element is written as 
\begin{equation}
ds^{2}=(a/\tau )^{2}(d\tau ^{2}-dr^{2}-r^{2}d\phi ^{2}-dz^{2}).  \label{ds2}
\end{equation}

We are interested in the local characteristics of quantum vacuum bilinear in
the field operator, such as, the vacuum expectation values (VEVs) of the
field squared and EMT. This quantities, as the expectation values of
products of the field operators and their derivatives at the same spacetime
point, formally diverge and a regularization procedure is required to
extract and renormalize the divergences. For a given field, those
divergences are completely determined by the local geometry of the
background spacetime (see, for example, \cite{Birr82,Park09}). In the
problems under consideration the local geometry in the region $r>0$ is the
same as that for dS spacetime and, hence, the divergences in the local
observables in that region are the same as well. The important conclusion
that follows from this is that the renormalization outside the string core
at $r=0$ is reduced to the one in dS spacetime. The VEVs in the latter
geometry are well investigated in the literature. Let us consider a local
observable of quantum vacuum given by the operator $F(x)$ at the spacetime
point $x=(\tau ,r,\phi ,z)$. For the geometry (\ref{ds2}) the corresponding
expectation value in the vacuum state $\left\vert 0\right\rangle $ can be
decomposed as%
\begin{equation}
\left\langle 0\right\vert F(x)\left\vert 0\right\rangle \equiv \left\langle
F(x)\right\rangle =\left\langle F(x)\right\rangle _{\mathrm{dS}%
}+\left\langle F(x)\right\rangle _{\mathrm{cs}}.  \label{Fdec}
\end{equation}%
Here, $\left\langle F(x)\right\rangle _{\mathrm{dS}}$ is the corresponding
VEV in dS spacetime when the cosmic string is absent and the contribution $%
\left\langle F(x)\right\rangle _{\mathrm{cs}}$ describes the effects induced
by the CS. In the region $r>0$ the latter is finite and the renormalization
is required for the first part only. The renormalization of different
physical observables in dS spacetime has been widely considered in the
literature and in the discussion below we are mainly interested in the
topological contribution $\left\langle F(x)\right\rangle _{\mathrm{cs}}$.

Another important point in quantum field theory in curved spacetime is the
choice of the vacuum state in the canonical quantization procedure. In
general, the vacuum state depends on the set of complete set of solutions to
the classical field equation used in the expansion of the field operator.
That is the case for the dS spacetime as well. Among different choices of
the vacuum state in dS background, the Bunch-Davies (BD) vacuum state \cite%
{Bunc78} is the most popular one. It shares the maximal symmetry of dS
spacetime. Among the set of one parameter family of vacuum states with
maximal symmetry \cite{Alle85} the BD vacuum is distinguished by the fact
that it is reduced to the Minkowski vacuum in the adiabatic limit and plays
a central role in inflationary scenario. In the discussion below it will be
assumed that the quantum fields are prepared in the BD vacuum state. The dS
spacetime and the BD vacuum are maximally symmetric and $\langle F(x)\rangle
_{\mathrm{dS}}$ in (\ref{Fdec}) does not depend on the spacetime point. As
it will be seen below, the string-induced part $\left\langle
F(x)\right\rangle _{\mathrm{cs}}$ depends on the radial and time coordinates
through the ratio $r_{\eta }=r/\eta $, where $\eta =|\tau |$. This property
is a consequence of the mentioned maximal symmetry in the absence of the CS.
Note that the proper distance $r_{p}$ from the string is given by the
combination $r_{p}=ar/\eta $ and, hence, $r_{\eta }$ is the proper distance
measured in units of the curvature radius $a$.

\subsection{VEV\ of the field squared for a scalar field}

We start our consideration with a scalar field $\varphi (x)$ with mass $m$
and curvature coupling parameter $\xi $. The field equation reads 
\begin{equation}
(\square +m^{2}+\xi R)\varphi (x)=0,  \label{Sceq}
\end{equation}%
where $\square =g^{ik}\nabla _{i}\nabla _{k}$ is the covariant d'Alembertian
operator and $R=12/a^{2}$ is the Ricci scalar in the region $r>0$. As vacuum
characteristics we will consider the VEVs of the field squared and EMT. In 
\cite{Beze09} the mode summation method has been used to evaluate the
Hadamard function for the BD vacuum state in the general case of $D$%
-dimensional dS spacetime. In this section, by using the results of \cite%
{Beze09}, we describe the properties of the scalar vacuum for the special
case $D=4$. The scalar mode functions $\varphi _{\sigma }(x)$ realizing the
BD state are specified by the set of quantum numbers $\sigma =(\lambda
,n,k_{z})$, where $0\leq \lambda <\infty $, $n=0,\pm 1,\pm 2,\ldots $, and $%
-\infty <k_{z}<+\infty $. They are expressed as 
\begin{equation}
\varphi _{\sigma }(x)=e^{i(\nu -\nu ^{\ast })\frac{\pi }{4}}\sqrt{\frac{%
q\lambda }{\pi }}\frac{\ \eta ^{3/2}}{4a}H_{\nu }^{(1)}(\gamma \eta
)J_{q|n|}(\lambda r)e^{in\phi +ik_{z}z},  \label{ModeSc}
\end{equation}%
where $\eta =|\tau |$, $\gamma =\sqrt{\lambda ^{2}+k_{z}^{2}}$, $q=2\pi
/\phi _{0}$, $H_{\nu }^{(1)}(u)$ and $J_{q|n|}(u)$ are the Hankel and Bessel
functions \cite{Abra72}, and 
\begin{equation}
\ \nu =\left( 9/4-12\xi -m^{2}a^{2}\right) ^{1/2}.  \label{nuSc}
\end{equation}%
Note that the parameter $\nu $ can be either non-negative real number or
purely imaginary complex number. In (\ref{ModeSc}), $\nu ^{\ast }$ stands
for the corresponding complex conjugate. We will assume that $\mathrm{Re\,}%
\nu <3/2$. For $\mathrm{Re\,}\nu \geq 3/2$ the BD vacuum state is not
physically realizable because of infrared divergences in the two-point
functions.

The topological contribution in the VEV of the field squared is presented in
the form 
\begin{equation}
\langle \varphi ^{2}\rangle _{\mathrm{cs}}=\frac{a^{-2}}{\sqrt{2}\pi ^{\frac{%
7}{2}}}\int_{0}^{\infty }dy\,\sqrt{y}e^{y-r_{\eta }^{2}y}K_{\nu
}(y)\int_{0}^{\infty }du\,g_{-}(q,u)K_{iu}(r_{\eta }^{2}y),  \label{phi2cs}
\end{equation}%
\newline
with $K_{\nu }(x)$ being the Macdonald function \cite{Abra72} and%
\begin{equation}
g_{\pm }(q,u)=\frac{1}{2}\left( e^{\pi u}\pm e^{-\pi u}\right) \left( \frac{1%
}{e^{2\pi u/q}\pm 1}-\frac{1}{e^{2\pi u}\pm 1}\right) .  \label{Fx}
\end{equation}%
The function $g_{+}(q,u)$ will be used below in the discussion of the
string-induced effects for the fermionc vacuum. For a conformally coupled ($%
\xi =1/6$) massless scalar field, $\nu =1/2$, the formula (\ref{phi2cs})
reduces to 
\begin{equation}
\langle \varphi ^{2}\rangle _{\mathrm{cs}}=\frac{q^{2}-1}{48\pi
^{2}a^{2}r_{\eta }^{2}}.  \label{phi2cc}
\end{equation}%
This formula is obtained from the corresponding result for cosmic string in
the Minkowski bulk (see below) replacing the distance $r$ by the proper
distance $r_{p}$ in dS spacetime. Of course, this corresponds to the
standard relation between two conformally related problems.

For points near the string we have $r_{\eta }\ll 1$ and the dominant
contribution to the integral in (\ref{phi2cs}) comes from large values of $y$%
. For those $y$ one has $K_{\nu }(y)\approx K_{1/2}(y)$ and the leading term
in the asymptotic expansion over $r_{\eta }$ is given by the right-hand side
of (\ref{phi2cc}). In the region $r_{\eta }\ll 1$ the dominant contribution
to the VEV comes from the fluctuations of wavelengths smaller than the
curvature radius and the influence of the gravitational field is weak. At
large distances from the CS the leading asymptotic is given by 
\begin{equation}
\langle \varphi ^{2}\rangle _{\mathrm{cs}}\approx \frac{(2r_{\eta })^{-3}}{%
\pi ^{D/2+1}a^{2}}\left\{ 
\begin{array}{ll}
(2r_{\eta })^{2\nu }C(q,\nu ), & \nu >0 \\ 
2|C(q,\nu )|\cos [2|\nu |\ln (2r_{\eta })+\psi _{1}], & \nu =i|\nu |%
\end{array}%
\right. ,  \label{phi2l}
\end{equation}%
where we have defined the function%
\begin{equation}
C(q,\nu )=\frac{\Gamma (\nu )}{\Gamma (2-\nu )}\int_{0}^{\infty
}du\,g_{-}(q,u)\Gamma \left( \frac{3}{2}-\nu +iu\right) \Gamma \left( \frac{3%
}{2}-\nu -iu\right) ,  \label{Cqunu}
\end{equation}%
and $\psi _{1}=\mathrm{arg\,}C(q,\nu )$. As seen, at large distances the
damping of the string-induced VEV is monotonic for real $\nu $ and
oscillatory for imaginary $\nu $.

\subsection{Energy-momentum tensor}

The expectation value $\langle T_{k}^{i}\rangle $ of the EMT is another
important local characteristic of the vacuum state. It is obtained from the
Hadamard function by acting the corresponding differential operator and then
taking the coincidence limit of the spacetime arguments. In the coordinates $%
x^{i}=(\tau ,r,\phi ,z)$, the nonzero components of the string-induced
contribution are expressed as (no summation over $k$):%
\begin{eqnarray}
\langle T_{k}^{k}\rangle _{\mathrm{cs}} &=&\frac{a^{-4}}{\sqrt{2}\pi ^{\frac{%
7}{2}}}\int_{0}^{\infty }du\,g_{-}(q,u)\ \int_{0}^{\infty }dy\,e^{-r_{\eta
}^{2}y}K_{iu}(r_{\eta }^{2}y)  \notag \\
&&\times \left\{ \left( 4\xi -1\right) \partial _{y}\left[ \left( 1-r_{\eta
}^{-2}\right) y\partial _{y}-\frac{3}{2}\right] -\frac{3\xi }{y}-\hat{g}%
_{k}(y)\right\} F_{\nu }(y),  \notag \\
\langle T_{0}^{1}\rangle _{\mathrm{cs}} &=&\frac{a^{-4}}{\sqrt{2}\pi ^{\frac{%
7}{2}}r_{\eta }}\ \int_{0}^{\infty }du\,g_{-}(q,u)\int_{0}^{\infty
}dy\,e^{-r_{\eta }^{2}y}K_{iu}(r_{\eta }^{2}y)\partial _{y}\left[ (4\xi
-1)y\partial _{y}+2\xi \right] F_{\nu }(y),  \label{Tmu}
\end{eqnarray}%
where we defined the function $F_{\nu }(y)=y^{\frac{3}{2}}e^{y}K_{\nu }(y)$
and the operators 
\begin{eqnarray}
\hat{g}_{0}(y) &=&(4\xi -1)\partial _{y}\left( y\partial _{y}\right) +1,\;%
\hat{g}_{3}(y)=-2\xi \partial _{y}+1,  \notag \\
\hat{g}_{1}(y) &=&(1-4\xi )r_{\eta }^{-2}\partial _{y}\left( y\partial
_{y}\right) -2\xi (1+r_{\eta }^{-2})\partial _{y}+1,  \label{g1} \\
\hat{g}_{2}(y) &=&r_{\eta }^{-2}\partial _{y}\left( y\partial _{y}\right)
-2\,\left[ \,y+\xi (1-r_{\eta }^{-2})\right] \partial _{y}-1.  \notag
\end{eqnarray}%
The off-diagonal component $\langle T_{0}^{1}\rangle _{\mathrm{cs}}$
corresponds to the energy flux density along the radial direction per unit
proper surface area. The energy flux is directed from the CS for $\langle
T_{0}^{1}\rangle _{\mathrm{cs}}>0$. Note that the energy flux from CS with
integer values of $q$ in some time-dependent background geometries has been
considered in \cite{Davi88} for massless scalar fields.

It can be checked that the trace relation $\langle T_{k}^{k}\rangle _{%
\mathrm{cs}}=\left[ 3(\xi -1/6)\square +m^{2}\right] \langle \varphi
^{2}\rangle _{\mathrm{cs}}$ is obeyed. In particular, for a conformally
coupled massless field the vacuum EMT is traceless. The conformal anomaly is
contained in the pure dS part $\langle T_{k}^{i}\rangle _{\mathrm{dS}}$. For
a conformally coupled massless scalar field the off-diagonal component $%
\langle T_{0}^{1}\rangle _{\mathrm{cs}}$ vanishes. In this case the vacuum
EMT is diagonal with the components 
\begin{equation}
\langle T_{k}^{i}\rangle _{\mathrm{cs}}=-\frac{q^{4}-1}{1440\pi
^{2}a^{4}r_{\eta }^{4}}\mathrm{diag}(1,1,-3,1),  \label{Tconf}
\end{equation}%
and the energy density is negative.

For points near the string, $r_{\eta }\ll 1$, the influence of the
gravitational field on the diagonal components of the EMT is week and the
leading order term reads (no summation over $k$)%
\begin{equation}
\langle T_{k}^{k}\rangle _{\mathrm{cs}}\approx \frac{B_{k}(\xi ,q)}{\left(
ar_{\eta }\right) ^{4}},\;B_{k}(\xi ,q)\equiv \left( 4\delta _{k2}-1\right) 
\frac{q^{2}-1}{144\pi ^{2}}\left[ d_{k}\left( 1-6\xi \right) +\frac{q^{2}+1}{%
10}\right] ,  \label{Tmunear}
\end{equation}%
where $d_{0}=d_{3}=-2$ and $d_{1}=d_{2}=1$. For a conformally coupled
massless scalar field (\ref{Tmunear}) coincides with the exact result (\ref%
{Tconf}). The off-diagonal component is a combined effect of the
gravitational field and CS and the corresponding asymptotic near the string
is given by%
\begin{equation}
\langle T_{0}^{1}\rangle _{\mathrm{s}}\approx \frac{q^{2}-1}{48\pi
^{2}a^{4}r_{\eta }^{3}}\left( 6\xi -1\right) .  \label{T01small}
\end{equation}%
This shows that in the case of minimally coupled field ($\xi =0$) and for
small $r_{\eta }$ the energy flux is directed towards the CS.

For real values of $\nu $, the large distance asymptotic ($r/\eta \gg 1$) is
given by (no summation over $k=1,2,3$)%
\begin{eqnarray}
\langle T_{0}^{0}\rangle _{\mathrm{cs}} &\approx &\left[ \frac{3-2\nu }{8}%
-\left( 2-\nu \right) \xi \right] \frac{6C(q,\nu )}{\pi ^{3}a^{4}\left(
2r_{\eta }\right) ^{3-2\nu }}\ ,  \notag \\
T_{k}^{k}\rangle _{\mathrm{cs}} &\approx &\frac{2\nu }{3}\langle
T_{0}^{0}\rangle _{\mathrm{cs}},\;\langle T_{0}^{1}\rangle _{\mathrm{cs}%
}\approx \frac{2\nu -3}{3r_{\eta }}\langle T_{0}^{0}\rangle _{\mathrm{cs}},
\label{Tkal}
\end{eqnarray}%
and we have a power-law decay as functions of the proper distance from the
CS. In the special case of conformal coupling the leading terms (\ref{Tkal})
vanish and the suppression of the topological contributions is stronger. For
imaginary $\nu $, similar to the case of the field squared, the decay of the
part $\langle T_{k}^{i}\rangle _{\mathrm{cs}}$ is damping oscillatory. The
diagonal components behave like $\sin [2|\nu |\ln (2r_{\eta })+\psi
_{1}^{(k)}]/r_{\eta }^{3}$ and for $\langle T_{0}^{1}\rangle _{\mathrm{cs}}$
the amplitude tends to zero as $1/r_{\eta }^{4}$.

\subsection{Minkowskian limit}

The geometry of CS in the background of Minkowski spacetime is obtained in
the limit $a\rightarrow \infty $ with fixed value of the synchronous time
coordinate $t$. In this limit one has $\eta \approx a-t$ and $\nu \approx ima
$. Introducing in (\ref{phi2cs}) a new integration variable $x=r_{\eta }^{2}y
$, we use the asymptotic estimate $K_{\nu }(y)\approx \sqrt{\pi /2y}%
e^{-m^{2}a^{2}/2y-y}$ for the Macdonald function with $y=xa^{2}/r^{2}$. For
the mean field squared we get%
\begin{equation}
\langle \varphi ^{2}\rangle _{\mathrm{cs}}^{\mathrm{(M)}}=\frac{1}{2\pi
^{3}r^{2}}\int_{0}^{\infty }dy\,e^{-y-\frac{m^{2}r^{2}}{2y}}\int_{0}^{\infty
}du\,g_{-}(q,u)K_{iu}(y).  \label{phi2M}
\end{equation}%
For a massless field the both integrals are evaluated by using the formulae
from \cite{Prud86} and one obtains%
\begin{equation}
\langle \varphi ^{2}\rangle _{\mathrm{cs}}^{\mathrm{(M)}}=\frac{q^{2}-1}{%
48\pi ^{2}r^{2}}.  \label{phi2M0}
\end{equation}%
In the Minkowski bulk the VEV of the field squared does not depend on the
curvature coupling parameter and this result could also be directly obtained
from (\ref{phi2cc}) in the limit $\eta \rightarrow a$.

The string induced part of the EMT is evaluated in the similar way and the
corresponding expression reads (no summation over $k$)%
\begin{equation}
\langle T_{k}^{k}\rangle _{\mathrm{cs}}^{\mathrm{(M)}}=-\frac{m^{4}}{8\pi
^{3}}\int_{0}^{\infty }dy\,\frac{e^{-\frac{m^{2}r^{2}}{2y}-y}}{y^{2}}%
G_{k}\left( \frac{2y}{m^{2}r^{2}}\right) \int_{0}^{\infty
}du\,g_{-}(q,u)K_{iu}(y),  \label{TmuM}
\end{equation}%
with the functions%
\begin{eqnarray}
G_{k}(x) &=&\left( 4\xi -1\right) \left( 1+x+x^{2}\right) +\frac{1}{2}%
m^{2}r^{2}x^{3},\;k=0,3,  \notag \\
G_{1}(x) &=&-2\xi x\left( 1+x\right) +\frac{1}{2}m^{2}r^{2}x^{3},  \notag \\
G_{2}(x) &=&2\xi \left( 2+3x+3x^{2}\right) -\,\,m^{2}r^{2}x^{2}\left( 1+%
\frac{3}{2}x\right) .  \label{TmuMG}
\end{eqnarray}%
Note that the relation $\langle T_{0}^{0}\rangle _{\mathrm{cs}}^{\mathrm{(M)}%
}=\langle T_{3}^{3}\rangle _{\mathrm{cs}}^{\mathrm{(M)}}$ is a consequence
of the Lorentz invariance with respect to boosts along the axis of the
cosmic string. The expressions (\ref{TmuM}) are further simplified for a
massless field with the result (no summation over $k$) \cite{Frol87,Dowk87} $%
\langle T_{k}^{k}\rangle _{\mathrm{cs}}^{\mathrm{(M)}}=B_{k}(\xi ,q)/r^{4}$.
For the case of a conformally coupled massless field this result is reduced
to the one obtained in \cite{Hell86}. For massive fields, integral
representations of the Green function are given in \cite{Line87}-\cite%
{More95}. The nonzero mass corrections to the VEVs of the field squared and
EMT in the limit $mr\ll 1$ were discussed in \cite{More95,Iell97}. The
closed expressions of those VEVs for the case of integer values of the
parameter $q$ are provided in \cite{Beze06b}. The vacuum energy for CS of
finite thickness is studied in \cite{Khus99}.

\section{Vacuum polarization for Dirac field}

\label{sec:Dirac}

In this section, based on \cite{Beze10}, we consider the CS polarization of
the BD vacuum for a massive 1/2-spin field.

\subsection{Fermionic modes}

The massive 1/2-spin field is described by a 4-component spinor $\psi (x)$
obeying the Dirac equation 
\begin{equation}
\left[ i\gamma ^{k}\left( \partial _{k}+\Gamma _{k}\right) -m\right] \psi
(x)=0,  \label{Direq}
\end{equation}%
with $4\times 4$ gamma matruces $\gamma ^{k}$ and spin connection $\Gamma
_{k}$. Here we use the representation of the Dirac matrices given by 
\begin{equation}
\gamma ^{0}=\left( 
\begin{array}{cc}
1 & 0 \\ 
0 & -1%
\end{array}%
\right) ,\;\gamma ^{l}=e^{-t/a}\left( 
\begin{array}{cc}
0 & \rho ^{l} \\ 
-\rho ^{l} & 0%
\end{array}%
\right) ,\;l=1,2,3.  \label{gaml}
\end{equation}%
The $2\times 2$ matrices $\rho ^{l}$ are the analogs of the Pauli matrices
in the cylindrical coordinate system: 
\begin{equation}
\rho ^{l}=\left( -\frac{i}{r}\right) ^{l-1}\left( 
\begin{array}{cc}
0 & e^{-iq\phi } \\ 
(-1)^{l-1}e^{iq\phi } & 0%
\end{array}%
\right) ,\;l=1,2,  \label{betal}
\end{equation}%
and $\rho ^{3}=\mathrm{diag}(1,-1)$.

The spinor eigenfunctions realizing the BD vacuum state are specified by the
set $\sigma =(\lambda ,j,k_{z},s=\pm 1)$ and are given by the expression 
\begin{equation}
\psi _{\sigma }(x)=\left( \frac{q\lambda e^{ma\pi }}{32\pi a^{3}}\right) ^{%
\frac{1}{2}}\eta ^{2}e^{iq(j-\frac{1}{2})\phi +ik_{z}z}\left( 
\begin{array}{c}
\sqrt{\gamma +sk_{z}}H_{1/2-ima}^{(1)}(\gamma \eta )J_{\beta _{j}}(\lambda r)
\\ 
\frac{si\lambda \epsilon _{j}}{\sqrt{\gamma +sk_{z}}}H_{1/2-ima}^{(1)}(%
\gamma \eta )J_{\beta _{j}+\epsilon _{j}}(\lambda r)e^{iq\phi } \\ 
-si\sqrt{\gamma +sk_{z}}H_{-1/2-ima}^{(1)}(\gamma \eta )J_{\beta
_{j}}(\lambda r) \\ 
\frac{\lambda \epsilon _{j}}{\sqrt{\gamma +sk_{z}}}H_{-1/2-ima}^{(1)}(\gamma
\eta )J_{\beta _{j}+\epsilon _{j}}(\lambda r)e^{iq\phi }%
\end{array}%
\right) ,  \label{psisig}
\end{equation}%
where $s=\pm 1$, $j=\pm 1/2,\pm 3/2,\ldots $, $\beta _{j}=q|j|-\epsilon
_{j}/2$, $\epsilon _{j}=\mathrm{sgn}(j)$, and the other notations are the
same as those for the scalar field. The mode functions (\ref{psisig}) are
the eigenfunctions of the projection $J_{3}$ of the total momentum: $\hat{J}%
_{3}\psi _{\sigma }(x)=qj\psi _{\sigma }(x)$. The two-point functions for
the Dirac field are obtained by summing the corresponding series over the
set of mode functions (\ref{psisig}) and their negative energy analogs.
Then, those functions are used for the evaluation of the fermion condensate
(FC) and the VEV of the EMT.

\subsection{Fermion condensate}

The FC is defined as the VEV $\langle 0|\bar{\psi}\psi |0\rangle \equiv
\langle \bar{\psi}\psi \rangle $, where $\bar{\psi}=\gamma ^{0}\psi
^{\dagger }$ is the Dirac adjoint. In the geometry at hand, the contribution
induced by the CS is expressed as%
\begin{equation}
\langle \bar{\psi}\psi \rangle _{\mathrm{cs}}=\frac{4\sqrt{2}}{\pi ^{\frac{7%
}{2}}a^{3}}\int_{0}^{\infty }dy\,y^{\frac{3}{2}}e^{y(1-r_{\eta }^{2})}\,{%
\mathrm{Im}}\left[ K_{\frac{1}{2}-im\alpha }(y)\right] \int_{0}^{\infty
}du\,g_{+}(q,u)\,{\mathrm{Im}}[K_{\frac{1}{2}-iu}(r_{\eta }^{2}y)],
\label{FCcs}
\end{equation}%
with the function $g_{+}(q,u)$ defined by (\ref{Fx}). It vanishes for a
massless field. The same is the case for the pure dS part $\langle \bar{\psi}%
\psi \rangle _{\mathrm{dS}}$ \cite{Beze10}. To clarify the behavior of the
FC as a function of the proper distance $r_{\eta }$ (in units of $a$) we
consider the asymptotic regions of small and large distances. The leading
term in the expansion near the CS, corresponding to $r_{\eta }\ll 1$, is
given by 
\begin{equation}
\langle \bar{\psi}\psi \rangle _{\mathrm{cs}}\approx \frac{m(q^{2}-1)}{%
24(\pi ar_{\eta })^{2}}.  \label{FCsmall}
\end{equation}%
The topological contribution is positive in that region and it dominates in
the total FC.

At large distances from the CS, $r_{\eta }\gg 1$, the influence of the
gravitational field is essential and the behavior of the topological FC is
described by 
\begin{equation}
\langle \bar{\psi}\psi \rangle _{\mathrm{cs}}\approx \frac{a|A(q,ma)|}{\pi
^{3}(ar_{\eta })^{4}}\,\sin \left[ 2ma\ln (2r_{\eta })-\varphi _{0}\right]
,\;r/\eta \gg 1,  \label{FClarge}
\end{equation}%
where $\varphi _{0}=\mathrm{arg\,}A(q,m\alpha )$ is the argument of the
complex function 
\begin{equation}
A(q,ma)=\frac{\Gamma (1/2-ima)}{\Gamma (5/2+ima)}\int_{0}^{\infty
}du\,ug_{+}(q,u)\Gamma \left( \frac{3}{2}+ima-iu\right) \Gamma \left( \frac{3%
}{2}+ima+iu\right) .  \label{AqRel}
\end{equation}%
At large distances we have damping oscillatory suppression with the
amplitude decaying as $1/r_{\eta }^{4}$. This behavior is in clear contrast
with the exponential decay in the geometry of CS in the Minkowski bulk. The
value of $r_{\eta }$ for the first zero of $\langle \bar{\psi}\psi \rangle _{%
\mathrm{cs}}$ increases with decreasing $ma$.

\subsection{EMT for Dirac field}

The topological contribution to the diagonal components of the VEV of the
EMT for Dirac field is presented in the form (no summation over $k$):%
\begin{equation}
\langle T_{k}^{k}\rangle _{\mathrm{cs}}=\frac{2^{\frac{3}{2}}}{\pi ^{\frac{7%
}{2}}a^{4}}\int_{0}^{\infty }dy\,y^{\frac{3}{2}}e^{y(1-r_{\eta }^{2})}\,\,{%
\mathrm{Re}}\left[ K_{\frac{1}{2}-ima}(y)\right] \int_{0}^{\infty
}du\,(2u)^{\delta _{k2}}g_{+}(q,u){\mathrm{Im}}[i^{\delta _{k2}}K_{\frac{1}{2%
}-iu}(r_{\eta }^{2}y)],  \label{Tlls}
\end{equation}%
and all the off-diagonal components vanish. Hence, unlike the case for a
scalar field, there is no energy flux in the fermionic vacuum. For the
diagonal components one has $\langle T_{0}^{0}\rangle _{\mathrm{cs}}=\langle
T_{1}^{1}\rangle _{\mathrm{cs}}=\langle T_{3}^{3}\rangle _{\mathrm{cs}}$. In
addition, the relation $\langle T_{k}^{k}\rangle _{\mathrm{dS}}=m\langle 
\bar{\psi}\psi \rangle _{\mathrm{dS}}$ for the trace of the EMT anf FC can
be checked. The massless fermionic field is conformally invariant and the
corresponding EMT $\langle T_{k}^{i}\rangle _{\mathrm{cs}}$ is traceless.

For massless fields the general formula is simplified to 
\begin{equation}
\langle T_{k}^{i}\rangle _{\mathrm{cs}}=-\frac{(q^{2}-1)(7q^{2}+17)}{2880\pi
^{2}a^{4}r_{\eta }^{4}}\mathrm{diag}(1,1,-3,1).  \label{Tklm0}
\end{equation}%
This expression is obtained from the VEV in the Minkowski bulk \cite%
{Frol87,Dowk87b} replacing the distance from the CS by the proper distance
in dS bulk. That is a consequence of the conformal relation between the
problems in the Minkowski and dS geometries for a massless Dirac field. In
the case of a massive fermionic field, the right-hand side of (\ref{Tklm0})
gives the leading term in the asymptotic expansion for small values of $%
r_{\eta }$. The asymptotic at large distances, corresponding to $r_{\eta
}\gg 1$, is obtained following the steps similar to the ones for the FC.
That gives (no summation over $k$) 
\begin{eqnarray}
\langle T_{k}^{k}\rangle _{\mathrm{cs}} &\approx &-\frac{|A(q,ma)|}{2\pi
^{3}a^{4}r_{\eta }^{4}}\cos \left[ 2ma\ln (2r_{\eta })-\varphi _{0}\right]
,\;k=0,1,3,  \notag \\
\langle T_{2}^{2}\rangle _{\mathrm{cs}} &\approx &\frac{|A(q,ma)|}{\pi
^{3}a^{4}r_{\eta }^{4}}\sqrt{9/4+m^{2}a^{2}}\sin [2ma\ln (2r_{\eta
})-\varphi _{0}+\varphi _{1}],  \label{T22l}
\end{eqnarray}%
where $\varphi _{1}=\arctan [3/(2m\alpha )]$. Again, we have an oscillating
decay at large distances.

\subsection{Fermionic vacuum densities in the Minkowski bulk}

The Minkowskian limit for the topological fermionic densities is obtained in
the way similar to that we have discussed for a scalar field. In the limit $%
a\rightarrow \infty $ and for fixed $x$, by using the estimate ${\mathrm{Im}}%
\left[ K_{1/2-im\alpha }(y)\right] \approx ma\sqrt{\pi }%
(2y)^{-3/2}e^{-m^{2}a^{2}/2y-y}$, with $y=xa^{2}/r^{2}$, for the FC in the
Minkowski bulk we get 
\begin{equation}
\langle \bar{\psi}\psi \rangle _{\mathrm{cs}}^{\mathrm{(M)}}=-\frac{2m}{\pi
^{3}r^{2}}\int_{0}^{\infty }dy\,e^{-\frac{m^{2}r^{2}}{2y}-y}\int_{0}^{\infty
}du\,g_{+}(q,u)\,{\mathrm{Im}}[K_{\frac{1}{2}-iu}(y)].  \label{FCM}
\end{equation}%
For the VEV of the EMT we use ${\mathrm{Re}}\left[ K_{1/2-ima}(y)\right]
\approx K_{ima}(y)$ and the asymptotic formula for $K_{ima}(y)$ given in the
previous section. In the Minkowskian limit that gives (no summation ove $k$)%
\begin{equation}
\langle T_{k}^{k}\rangle _{\mathrm{cs}}^{\mathrm{(M)}}=\frac{2}{\pi ^{3}r^{4}%
}\int_{0}^{\infty }dx\,x\,e^{-\frac{m^{2}r^{2}}{2x}-x}\int_{0}^{\infty
}du\,(2u)^{\delta _{k2}}g_{+}(q,u){\mathrm{Im}}[i^{\delta _{k2}}K_{\frac{1}{2%
}-iu}(x)].  \label{TkM}
\end{equation}%
The expressions (\ref{FCM}) and (\ref{TkM}) coincide with the formulas found
in \cite{Beze08} (the misprint for the component $k=0$ is corrected). For a
massless field the FC vanishes and the vacuum EMT in the Minkowski bulk \cite%
{Frol87,Dowk87b} is obtained from (\ref{Tklm0}) taking $\eta =a$.
Alternative representations for the CS induced FC and EMT for massive spinor
fields on the Minkowski bulk are given in \cite{Beze06,Bell14}. The effects
of a finite radius magnetic flux are discussed in \cite{Maio17}.

\section{Electromagnetic field}

\label{sec:Elmag}

Now we turn to the case of the electromagnetic field. For the VEVs of the
electric and magnetic fields squared one has the decomposition (\ref{Fdec})
with the topological parts%
\begin{equation}
\langle E^{2}\rangle _{\mathrm{cs}}=\langle B^{2}\rangle _{\mathrm{cs}}=-%
\frac{\left( q^{2}-1\right) \left( q^{2}+11\right) }{180\pi a^{4}r_{\eta
}^{4}}.  \label{E21}
\end{equation}%
The expression for the VEV of the energy-momentum tensor reads%
\begin{equation}
\langle T_{k}^{i}\rangle _{\mathrm{cs}}=-\frac{\left( q^{2}-1\right) \left(
q^{2}+11\right) }{720\pi ^{2}a^{4}r_{\eta }^{4}}\,\mathrm{diag}(1,1,-3,1).
\label{Tel}
\end{equation}%
In 4-dimensional spacetime the electromagnetic field is comformally
invariant and the VEVs (\ref{E21}) and (\ref{Tel}) are obtained from the
VEVs in the Minkowski bulk \cite{Frol87,Dowk87b} by the replacement $%
r\rightarrow ar_{\eta }$.

In the spacetime dimensions $D$ different from 4 the electromagnetic field
is not conformally invariant and a simple conformal relation between the
VEVs in the locally Minkowski and dS bulks does not take place. The
electromagnetic vacuum densities for general case of $D$ have been
considered in \cite{Saha17,Saha23}. Relatively compact expressions are
obtained for even values of $D$. For example, in the case $D=6$ one has%
\begin{eqnarray}
\langle E^{2}\rangle _{\mathrm{cs}} &=&-\left( q^{2}-1\right) \frac{%
q^{4}+22q^{2}+211}{1890\pi ^{2}a^{6}r_{\eta }^{6}},  \notag \\
\langle B^{2}\rangle _{\mathrm{cs}} &=&-\left( q^{2}-1\right) \frac{%
2q^{4}+23q^{2}+191-21\left( 3+r_{\eta }^{2}\right) \left( q^{2}+11\right) }{%
3780\pi ^{2}a^{6}r_{\eta }^{6}}.  \label{B2D6}
\end{eqnarray}%
At large distances one has $|\langle B^{2}\rangle _{\mathrm{cs}}|\gg
|\langle E^{2}\rangle _{\mathrm{cs}}|$. In dimensions $D\neq 4$ the
energy-momentum tensor has a nonzero off-diagonal component $\langle
T_{0}^{1}\rangle _{\mathrm{cs}}$ describing the energy flux along the radial
direction \cite{Saha23}. In particular, for $D=6$ one has%
\begin{equation}
\langle T_{0}^{1}\rangle _{\mathrm{cs}}=-\frac{\left( q^{2}-1\right)
(q^{2}+11)}{720\pi ^{3}a^{6}r_{\eta }^{5}}.  \label{T01D6}
\end{equation}%
The corresponding energy flux is directed towards the cosmic string. 

\section{Summary}

We have discussed the vacuum polarization effects by CS in dS spacetime. As
local characteristics of the quantum vacuum the expectation values of the
field squared and EMT are studied. For spinor fields the FC is considered as
well. For points near the CS the influence of the gravitational field on the
field squared and on the diagonal components of the EMT is week and the
leading terms in the corresponding asymptotic expansions are obtained from
those for CS in the Minkowski bulk replacing the distance from the string by
the proper distance in dS spacetime. The effects of the spacetime curvature
are essential at proper distances of the order of or larger than the dS
curvature radius. In particular, at large distances the fall-off of the
topological contributions, as functions of the proper distance, follows
power-law instead of the exponential decay in the Minkowski bulk for massive
fields. For scalar fields, depending on the mass, two regimes are realized
with monotonic or oscillatory decrease. For spinor fields the fall-off is
always damping oscillatory, whereas for the electromagnetic field the VEVs
follow monotonic power-law decay. A qualitatively new feature for CS in dS
bulk is the presence of nonzero off-diagonal component $\langle
T_{0}^{1}\rangle _{\mathrm{cs}}$ for scalar (except the case of a
conformally coupled massless field) and the electromagnetic (in spacetime
dimensions $D\neq 4$) fields. It describes the energy flux in the vacuum
state along the radial direction.

\section*{Acknowledgments}

The work was supported by the grants No. 21AG-1C047 and 24FP-3B021 of the
Higher Education and Science Committee of the Ministry of Education,
Science, Culture and Sport RA.

\end{document}